\documentstyle[aps,epsf]{revtex}
\begin{document}

\title{Modified Zimanyi-Moszkowski Model for Finite Nuclei}
\author{A. Delfino$^a$, M. Chiapparini$^b$ and W. A. Colchete$^a$}
\address{$^a$Instituto de F\'{\i}sica - Universidade Federal
Fluminense \\
Av. Litor\^anea s/n , \,24210-340 Boa Viagem, Niter\'oi \\
Rio de Janeiro, Brazil}
\address{$^b$Instituto de F\'{\i}sica, Universidade do Estado do
Rio de Janeiro,\\
20559-900, Maracan\~a, Rio de Janeiro, RJ, Brazil.}
\vspace{.3 cm}
\date{\today}
\maketitle

\begin{abstract}
      By scaling the mesonic masses in the Walecka model, to the
      inverse of the hadronic effective mass in medium, we generate
      the Zimanyi-Moszkowski (ZM) models added to a surface term.
      For infinite nuclear matter, the surface term vanishes and this
      new model becomes equivalent to the usual ZM model. For
      finite nuclei calculations, the surface contribution changes
      the spin-orbit splitting in the right experimental direction.
      Calculations for some nuclei are presented.
\end{abstract}


\section{Introduction}

The relativistic linear $\sigma-\omega$ model
(hereafter called Walecka model) \cite{walecka}
satisfactorily explains many properties of nuclear
matter and finite nuclei.
A shortcoming of this model is, however, the prediction of a high value for the
compression modulus $K=550$ MeV.
The introduction of nonlinear scalar
self-coupling terms \cite{boguta} has brought $K$ to a reasonable value of
$250$ MeV in a theory with four free parameters. Modifications of this
kind of model have been implemented by many authors \cite{miya,furns}. Zimanyi
and Moszkowski (ZM)  \cite{zm} and Heide and Hudaz \cite{heide},
aiming to keep only two free parameters have proposed nonlinear models,
obtaining soft equations of state. The results for the compression
modulus, $K=224$ MeV, and nucleonic effective mass, $M^\ast =797$ MeV, compare
very well with Skyrme-type calculations \cite{z1}.
Many successful applications
of ZM model have been done since its original proposal, regarding
for example, quantum molecular dynamics approach \cite{feld1}, neutron
stars \cite{glend1}, quark and gluon condensates in medium \cite{delf3}.
In these applications, the softness of ZM  model
is essential for the obtaining of a desirable behavior at high
density regimes. On the other hand, however, finite nuclei
calculations showed that the spin-orbit interaction is too small to explain
the observed spin-orbit splitting for finite nuclei \cite{sharma,wald,koepf,chiappa}.

Walecka and ZM models became therefore the extreme
of the most simple quantum-hadron-dynamics models. The
first with too much relativistic content while the second with
too little. Both models are very simple and only differ in the coupling
among the fields.

Recently Bir\'o and Zimanyi \cite{bz} proposed a new effective
Lagrangian, adding to the usual ZM-Lagrangian a tensor coupling
analogous to the one which leads to the anomalous gyromagnetic
ratio. An additional free parameter in this term
is suggested to be eliminated in favor of the improvement of
the spin-orbit splitting for finite nuclei calculations. In this
work, we intend to exhaust first the possibilities of the ZM model,
still in a two-free parameters version, to improve the
spin-orbit splitting for finite nuclei calculations. In Sec. II
we show how we scale the mesonic masses in the Walecka model,
in such a way that we generate
the original ZM model added of a surface term,
which does not contribute for infinite nuclear matter
but changes the results for finite nuclei calculations,
presented in Sec. III.

\section{Modified ZM Model}

Let us start connecting the simple Walecka and ZM models,
through an unified  Model Lagrangian density \cite{delf1}

\begin{equation}
      {\cal L}_{M}=\bar\psi \left\{
      \gamma^\mu (i\partial_\mu-g_\omega\omega_\mu)
      -f(\sigma)M \right\}\psi
      - \frac{1}{4} \omega^{\mu\nu} \omega_{\mu\nu}
      + \frac{1}{2} m^2_\omega\omega_\mu\omega^\mu
      + \frac{1}{2} \left( \partial_\mu\sigma
      \partial^\mu\sigma - m^2_\sigma\sigma^2 \right) \; ,
      \label{1}
\end{equation}

where the degrees of
freedom are the baryon field $\psi$, the scalar meson field $\sigma$
and the vector meson field $\omega^\mu$. The real function $f(\sigma)$ is to
be defined according with each model under consideration, with the condition
that for zero density (it means, $\sigma$ going to zero) $f(\sigma)$ goes to
one, and for higher densities
the effective baryonic mass must approximate to zero asymptotically. Of course,
$f(\sigma)$ also specifies the kind of scalar meson-nucleon coupling.
The Dirac equation obtained from the Lagrangian density (\ref{1}) gives
$ f(\sigma)=M^\ast/M\equiv m^\ast$ where $M$ and $M^\ast$ are the bare and
in-medium effective baryonic mass respectively (hereafter we will interpret
$\ast$ as referring to effective quantities in the medium.)

The Walecka model can be obtained as a particular case of the model defined by
${\cal L}_M$ making the choice
$f(\sigma)=(1-g_\sigma\sigma/M)$, while the usual ZM
model is obtained by the choice
$f(\sigma)=(1+g_\sigma\sigma/M)^{-1}$.
It is clear now that a connection between both models can be
obtained if one redefines the scalar coupling constant in the
Walecka model. For short:

\begin{equation}
      {\cal L}_{\rm ZM} \equiv {\cal L}_{Walecka}(g_\sigma \rightarrow g_\sigma
      ^\ast)\; , \label{2}
\end{equation}

where $g_\sigma^\ast$ is now a function of $\sigma$, given by

\begin{equation}
      g_\sigma^\ast=g_\sigma f_{\rm ZM}(\sigma)=
      g_\sigma m^\ast = g_\sigma (1 + g_\sigma\sigma/M)^{-1}
      \; . \label{3}
\end{equation}

As shown in Ref. \cite{delf1}, a modified version of the usual ZM model
(called ZM3 in Ref. \cite{delf2}) may be obtained from the Walecka
model by performing a redefinition in both mesonic coupling constants,

\begin{equation}
      {\cal L}_{\rm ZM3} \equiv {\cal L}_{Walecka}
      (g_\sigma\rightarrow g_\sigma^{\ast}\:;\: g_\omega\rightarrow
      g_\omega^\ast)\;, \label{4}
\end{equation}

where

\begin{equation}
      \frac{g_\sigma^\ast}{g_\sigma} =
      \frac{g_\omega^\ast}{g_\omega} = m^\ast \; . \label{5}
\end{equation}

Note that Eqs. (\ref{2}) and (\ref{4}) simplify the understanding of
different kinds of ZM models since they can now be understood
as directly coming from the Walecka model  where the coupling
constants become density dependent.

Now we pose the question whether there is another connection
among these models through a rescaling of the mesonic masses.
This question
arises  quite naturally once we know that for Walecka model as well
as for ZM models what matters for the saturation of the infinite
nuclear matter are the ratios
$C_\sigma^2 = g_\sigma^2M^2/m_\sigma^2$ and
$C_\omega^2 = g_\omega^2M^2/m_\omega^2$. To answer this
question we start with the following Lagrangian density

\begin{eqnarray}
      {\cal L}^\prime_{W}&=&\bar\psi \left\{
      \gamma^\mu (i\partial_\mu-g_\omega\omega_\mu)
      -(M-g_\sigma \sigma) \right\}\psi  \nonumber \\
      &&
      - \frac{1}{4} \omega^{\mu\nu} \omega_{\mu\nu}
      + \frac{1}{2} m^2_\omega\omega_\mu\omega^\mu
      + \frac{1}{2} \left( \partial_\mu\sigma
      \partial^\mu\sigma - {m^\ast}^2_\sigma\sigma^2 \right) \; .
      \label{6}
\end{eqnarray}

After performing the rescaling $\sigma \rightarrow h(\sigma )\sigma$
and imposing $m^\ast_\sigma=m_\sigma/h(\sigma)$, with
$h(\sigma)=(1-g_\sigma\sigma /M)$, we get

\begin{eqnarray}
      {\cal L}_{\rm MZM}&=&\bar\psi \left\{
      \gamma^\mu (i\partial_\mu-g_\omega\omega_\mu)
      -M(1+g_\sigma\sigma/M)^{-1} \right\}\psi
      - \frac{1}{4} \omega^{\mu\nu} \omega_{\mu\nu}
      + \frac{1}{2} m^2_\omega\omega_\mu\omega^\mu  \nonumber \\
      &&
      - \frac{1}{2} m_\sigma\sigma^2
      + \frac{1}{2}\frac{1}{\left(1-g_\sigma\sigma/M\right)^4}
        \partial_\mu\sigma\partial^\mu\sigma \; .
      \label{7}
\end{eqnarray}

We refer to this Lagrangian as ${\cal L}_{\rm MZM}$, since
at the level of Mean Field Approximation (MFA),  where
the derivative mesonic terms do not contribute, it becomes the
usual ZM Lagrangian. Shortly, the connection between the Walecka
model and this modified ZM model is

\begin{equation}
      {\cal L}_{\rm MZM} \equiv {\cal L}_{Walecka}
      (m_\sigma\rightarrow m_\sigma^\ast)\; , \label{8}
\end{equation}

and

\begin{equation}
      {\cal L}_{\rm MZM3} \equiv {\cal L}_{Walecka}
      (m_\sigma\rightarrow m_\sigma^\ast \: ; \:
       m_\omega\rightarrow m_\omega^\ast )
      \; ,\label{9}
\end{equation}

where

\begin{equation}
      m^\ast_\sigma = \frac{m_\sigma}{1-g_\sigma\sigma/M} \; ,
      \label{10}
\end{equation}

and

\begin{equation}
      m^\ast_\omega = \frac{m_\omega}{1-\alpha g_\sigma\sigma/M} \; ,
      \label{11}
\end{equation}

with $\alpha = 0$ for the MZM model and $\alpha = 1$ for the MZM3 model.

In this model the Euler-Lagrange equations can be written as

\begin{equation}
      \left[\gamma^\mu (i \partial_\mu - g_\omega\omega_\mu )
      - (M-g_\sigma\sigma)\right]\psi=0 \; , \label{12a}
\end{equation}

\begin{equation}
      \partial_\mu \omega^{\mu\nu} + {m^\ast}^2_\omega\omega^\nu =
      g_\omega \bar{\psi}\gamma^\nu \psi \; , \label{12b}
\end{equation}

\begin{equation}
      \partial_\mu \partial^\mu \sigma +
      \frac{{m^\ast}^3_\sigma}{m_\sigma}\sigma =
      g_\sigma \bar{\psi}\psi +
      \alpha \frac{g_\sigma}{M}\frac{{m^\ast}^3_\omega}{m_\omega}
      \omega_\mu\omega^\mu \; . \label{12c}
\end{equation}

Rescaling the mesonic fields in the form
$\sigma = (1+g_\sigma\sigma^\prime/M)\sigma^\prime$ and
$\omega^\mu = (1+\alpha g_\sigma\sigma^\prime/M){\omega^\prime}^\mu$,
with the scalar fiels related through $1+g_\sigma\sigma^\prime/M
=(1-g_\sigma\sigma/M)^{-1}$ we obtain, at the MFA level, the equations of
motion of the ZM models \cite{chiappa}.

Eqs. (\ref{2}) and (\ref{4}) as well as Eqs. (\ref{8}) and (\ref{9}) indicate
how to obtain the ZM models from the Walecka model within the MFA.
The last set of relations, Eqs. (\ref{8}) and (\ref{9}), is particularly interesting
by the following. First, the usual ZM model written in this form, becomes now
clearly a particular case of the nonlinear Walecka model \cite{boguta}.
Indeed, we have expanded Eq. (\ref{10}) up to order $\sigma^2$, what
means to get a nonlinear Walecka model with scalar cubic and quartic
terms, and observed that the changes in the nuclear matter bulk properties
are not more than a few percent \cite{wagner}.
Second, the scaling exhibited by Eq. (\ref{10}) and also by Eq. (\ref{11})
points exactly to the inverse of the Brown-Rho scaling \cite{brown-rho},
obtained from chiral model Lagrangians. Brown-Rho
scaling claims that in the medium the mesonic masses  should scale
as $m^\ast$. ZM models, however, can be seen now as hadronic
models where mesonic masses scale as $1/m^\ast$.
Third, the connection is only true for the infinite nuclear
matter where the last term of Eq. (\ref{7}) is identically zero in the MFA.
For finite nuclei surface effects
are important, even in MFA, and this last term has to be considered anyway,
changing the known results for ZM models.

\section{Results}

The nuclear matter set of parameters
$C_\sigma^2 = g_\sigma^2M^2/m_\sigma^2$,
$C_\omega^2 = g_\omega^2M^2/m_\omega^2$ and
$C_\rho^2 = g_\rho^2M^2/m_\rho^2$
for MZM and MZM3 are the same of
ZM and ZM3 presented in Ref. \cite{chiappa}.
Also the same as ZM and ZM3
are the incompressibility $K$, the nucleonic
effective mass $m^\ast$, the scalar and vector potentials, $S$ and $V$,
given in Table \ref{t1}. This happens because, for infinite nuclear matter,
the last term of Eq. (\ref{7}) does not contribute.

\begin{figure}
\epsffile{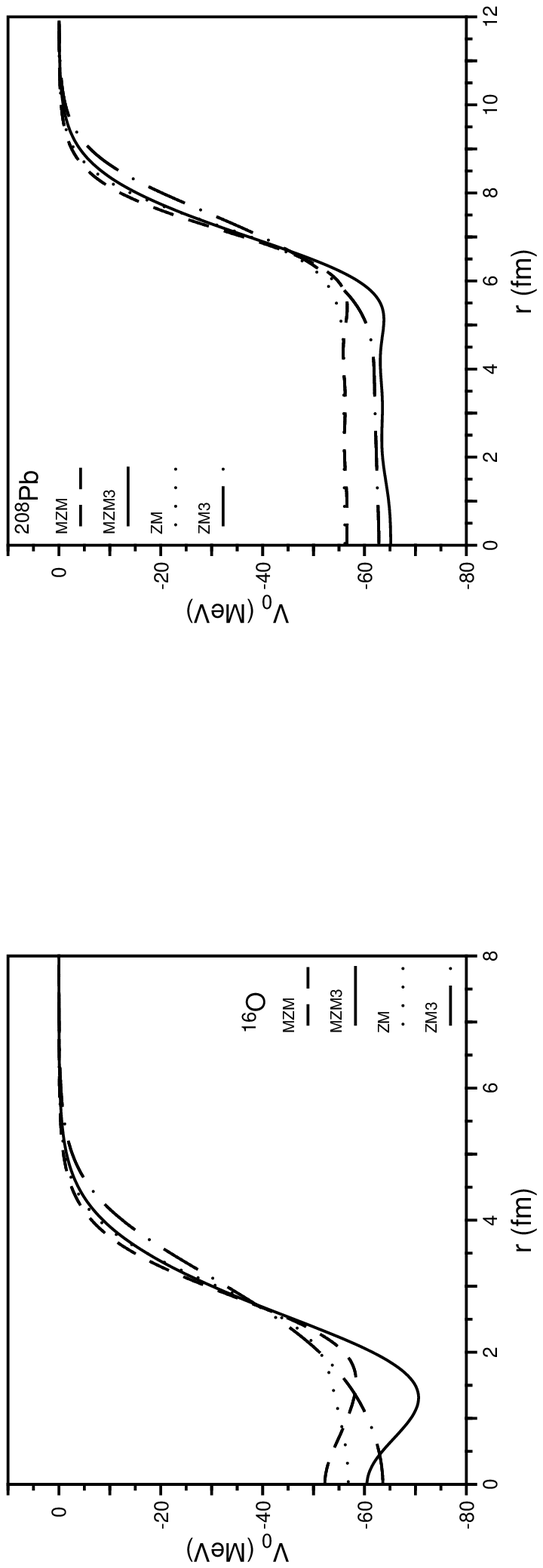}
\caption{The central potential $V_0$ for $^{16}$O and $^{208}$Pb nuclei.
Besides the curves for the models MZM-MZM3 presented in this work, the
usual ZM-ZM3 ones are shown.}
\label{f1}
\end{figure}

\begin{figure}
\epsffile{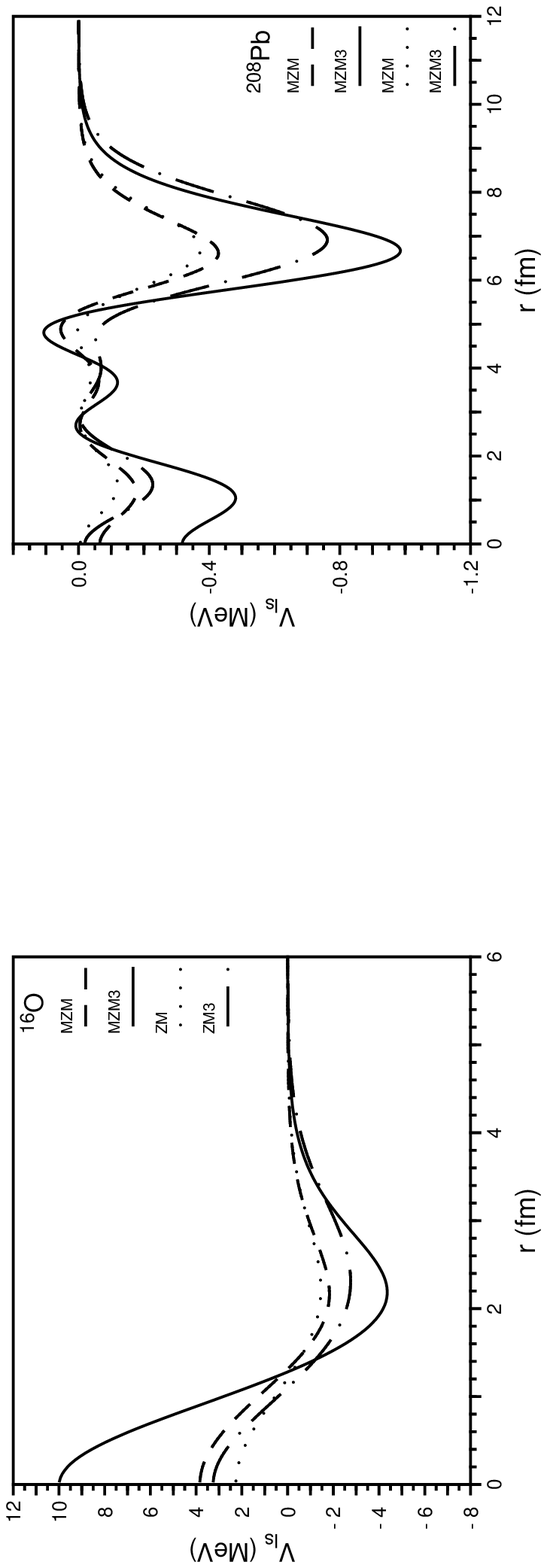}
\caption{The spin-orbit potential $V_{ls}$ for $^{16}$O and $^{208}$Pb nuclei.
Besides the curves for the models MZM-MZM3 presented in this work, the
usual ZM-ZM3 ones are shown.}
\label{f2}
\end{figure}

The calculations for finite nuclei follows the steps
we have presented in Ref. \cite{chiappa}. Tables \ref{t2}-\ref{t3} show the
results
obtained for the static ground-state properties in $^{16}$O and $^{208}$Pb
One feature to point out is that the modified versions
predict a {\em r.m.s.} for the charge radius
that is slightly smaller than the one calculated with the
old version. For the binding energy this trend reverts itself.
Consistent with a expected surface term effect, the last term of
Eq. (\ref{7}) is more relevant for $^{16}$O  (from the shell model we know that
most of the nucleons are in the surface), decreasing as the
nuclei becomes heavier and giving a nearly nuclear matter behavior as,
for example, in the $^{208}$Pb nucleus. The systematic
behavior for the {\em r.m.s.} and the binding energy
is due to the slightly deeper central potential presented in
Fig. \ref{f1} for $^{16}$O and $^{208}$Pb nuclei. Regarding the
spin-orbit splitting, the systematic may be understood through
Fig. \ref{f2}. In Tables \ref{t4}-\ref{t5} we
present the spectra for the $^{16}$O and $^{208}$Pb nuclei,
for MZM and MZM3 compared with the old
ZM and ZM3 versions. Note how the $\Delta\varepsilon_{ls}$ for
$^{16}$O differs more than that for $^{208}$Pb, between the old
and the new version models. The contributions for $^{40}$Ca,
$^{48}$Ca and $^{90}$Zr from the surface term in
MZM and MZM3 lie in between the curves presented in
Fig. \ref{f1} \cite{wagner}.

We have presented two distinct ways to arrive at the
ZM model starting from the Walecka model. The first, already presented in
Ref. \cite{chiappa}, consists in the rescaling of the mesonic coupling
constants as described by Eqs. (\ref{2})-(\ref{5}). In this case the
nucleonic effective mass $m^\ast$ is defined as Eq. (\ref{3}), in agreement with
the definition of effective mass in the ZM model. In this case,
the ZM model is reobtained obtained for infinite nuclear matter and for finite nuclei
as well.  The second one, in
which consists the main contribution of this work, where ZM model is
obtained from the Walecka model by rescaling the mesonic masses
as given by Eqs. (\ref{8})-(\ref{11}). The identification is clear only for
infinite nuclear matter, since a surface term (the last term of
Eq. (\ref{7})) remains for finite nuclei calculation.
In this case the
nucleonic effective mass is defined as $m^\ast=1-g_\sigma\sigma/M$,
in agreement  with the definition of effective mass in the Walecka model.
This effective local mass is shown in Fig. \ref{3} for the $^{16}$O and
$^{208}$Pb
nuclei, where is clear the surface enhanced behavior of the modified models.
The modified versions MZM and MZM3 change the spin-orbit splitting  in the
experimental
direction, as we can see from the Tables \ref{t2}-\ref{t3}. In the particular
case of $^{16}$O,
where the fail of ZM and ZM3 was more visible (see Ref.
\cite{chiappa}), the new version MZM3 brings the
 $p_{3/2} - p_{1/2}$ spin-orbit splitting to 4 MeV in a model with two free
parameters. If one intends to improve the ZM models,
following the interesting suggestion of Bir\'o and Zimanyi \cite{bz},
the presented modified versions MZM and MZM3 may be
a better option for the starting point than the usual ZM models, since
they provide already nearly experimental spectra for finite nuclei. Still
thinking about a hadronic model with few parameters (three for example)
one could use $\alpha$ in Eq. (\ref{11}) as a free parameter to fit the
spin-orbit splitting of $^{16}$O. The purpose of this work is however, by
keeping the two varying parameters under control, {\it i)} to show the
possibles ways from where the ZM models came from,
{\it ii)} to attempt for the surfer contribution in finite nuclei, when mesonic
scaling mass takes place and
{\it iii)} to stats that ZM models incorporate a mesonic mass scaling to the
inverse of the effective baryonic mass. By design this, we can understand
simple models before to start increasing the number of free parameters to
improve the observables, {\it i.e.}, to have the major parameters under control.

Summarizing, we have presented a new discussion on
the ZM models. Our main conclusions are as follows:

The ZM models \cite{zm} have been presented in their
original version as coming from the inclusion of
a derivative coupling into the original Walecka model.
Further, it was established that it is equivalent to a
linear scaling of the coupling constants of the Walecka
model with the effective nucleonic mass
$m^\ast=1/(1+g_{\sigma}\sigma/M)$ of the ZM model \cite{delf1}. Here, in
another kind of equivalence, ZM models may be seen as
also coming from the rescaling of the mesonic masses
in the Walecka model to the inverse of the nucleonic
effective mass $m^\ast=1-g_\sigma\sigma/M$.

This last equivalence exactly applies to nuclear
matter, since in  MFA the derivative mesonic terms
do not contribute. However, for finite nuclei,
surface terms become important and the usual ZM models
lead quite naturally to be modified (MZM), without
any new free parameter and having the same features
of the nuclear matter ZM models.

We have performed calculations with MZM models
for several finite nuclei and compared the results with
the usual ZM models. We see that the spin-orbit splitting,
usually a drawback in the ZM models, changes in the
right experimental direction, when calculated with
the MZM models. Consistent with the theory,
we observe that the surface term contribution present in
MZM models decreases as the size of the nuclei increases.

Recently, many authors \cite{saito1} have addressed the question
that since the nucleon is not a point object, but has
structure, it should afford changes when inside the
nuclear medium. In this context, it is strongly
conjectured that the mesonic coupling constant
should be density dependent. However, we believe that
there is no especial reason why the mesonic masses
could not also become effective, changing in the
nuclear medium.  Indeed, an effective density mesonic mass
dependence has been conjectured in the analysis of the
naturalness in the quark-meson coupling model \cite{saito2}.
In this sense, our work provides a contribution to a
better understanding of the medium-dependent mesonic
coupling and mass parameters of hadronic models.

\vspace{1cm}

{\bf Acknowledgements}

A. D. thanks to CNPq and M. C. thanks to FAPERJ for partial financial
support.

\begin{table}
      \begin{tabular}{ccccc}
            Model   &  $m^*$  &  $\kappa $(MeV)  & $S$ (MeV) &  $ V$ (MeV) \\ \hline
            ZM      &   0.85  &    225    &   -141    &    82    \\
            ZM3     &   0.72  &    156    &   -267    &    204
      \end{tabular}
      \caption{The incompressibility $K$, the nucleonic
      effective mass $m^\ast$, and the scalar and vector
      potentials, $S$ and $V$, in nuclear matter for the usual ZM models.}
      \label{t1}
\end{table}

\begin{table}
      \begin{tabular}{cccccc}
                               &ZM  &MZM &ZM3 &MZM3&Exp. \\ \hline
            $\varepsilon$ (MeV)&8.40&9.37&7.50&9.42&7.98\\
            $\langle r^2 \rangle$ (fm$^2$)&2.64&2.56&2.78&2.56&2.74\\
            $\varepsilon_{1p1/2}-\varepsilon_{1p3/2}$ (MeV)&
            1.4(1.4)&1.6(1.6)&2.9(2.9)&4.0(4.0)&6.1(6.3)
      \end{tabular}
      \caption{Binding energy, mean squared charge radius and spin-orbit
       splitting for the $^{16}$O nucleus. Values between parenthesis are
       for protons, the others are for neutrons.}
      \label{t2}
\end{table}

\begin{table}
      \begin{tabular}{cccccc}
                               &ZM  &MZM &ZM3 &MZM3&Exp. \\ \hline
            $\varepsilon$ (MeV)&7.86&8.17&7.66&8.28&7.87\\
            $\langle r^2 \rangle$ (fm$^2$)&5.54&5.51&5.66&5.52&5.50\\
            $\varepsilon_{2p1/2}-\varepsilon_{2p3/2}$ (MeV)&
            0.2(0.2)&0.2(0.2)&0.4(0.5)&0.5(0.6)&0.5\\
            $\varepsilon_{2f7/2}-\varepsilon_{2f5/2}$ (MeV)&
            0.5&0.6&1.2&1.3&1.8\\
            $\varepsilon_{3p1/2}-\varepsilon_{3p3/2}$ (MeV)&
            0.2&0.2&0.4&0.5&0.9
      \end{tabular}
      \caption{Binding energy, mean squared charge radius and spin-orbit
       splitting for the $^{208}$Pb nucleus. Values between parenthesis are
       for protons, the others are for neutrons.}
      \label{t3}
\end{table}

\begin{table}
      \begin{tabular}{cccccc}
             Level &ZM  &MZM &ZM3 &MZM3&Exp. \\ \hline
            $1s_{1/2}$&35.4(31.2)&37.7(33.4)&
             36.2(32.1)&43.2(38.7)&47.0(40$\pm$8)\\
            $1p_{3/2}$&19.6(15.6)&21.1(17.0)&
             19.4(15.6)&23.3(19.2)&21.8(18.4)\\
            $1p_{1/2}$&18.2(14.2)&19.5(15.4)&
             16.5(12.7)&19.4(15.2)&15.7(12.1)
      \end{tabular}
      \caption{Energy spectra for the $^{16}$O nucleus. Energies are in MeV.
      Values between parenthesis are for protons, the others are for neutrons.}
      \label{t4}
\end{table}

\begin{table}
      \begin{tabular}{llllll}
             Level &ZM  &MZM &ZM3 &MZM3&Exp. \\ \hline
            $1s_{1/2}$&45.1(35.6)&45.4(36.2)&
              50.6(40.0)&52.3(42.0)&\\
            $1p_{3/2}$&41.0(31.8)&41.4(32.6)&
              45.3(35.2)&47.3(37.5)&\\
            $1p_{1/2}$&40.7(31.6)&41.3(32.5)&
              45.1(34.9)&47.0(37.2)&\\
            $1d_{5/2}$&35.8(26.9)&36.5(27.9)&
              39.0(29.4)&41.1(31.8)&\\
            $1d_{3/2}$&35.6(26.6)&36.2(27.6)&
              38.4(28.7)&40.5(31.1)&\\
            $1f_{7/2}$&29.9(21.2)&30.6(22.2)&
              31.9(22.7)&34.0(25.0)&\\
            $1f_{5/2}$&29.4(20.6)&30.1(21.6)&
              30.9(21.5)&32.9(23.8)&\\
            $1g_{9/2}$&23.2(14.6)&23.9(15.6)&
              24.3(15.4)&26.2(17.5)&\\
            $1g_{7/2}$&22.4(13.8)&23.1(14.8)&
              22.6(13.6)&24.4(15.6)&(11.4)\\
           $1h_{11/2}$&15.9(7.41)&16.6(8.27)&
              16.2(7.65)&17.7(9.24)&(9.4)\\
            $1h_{9/2}$&14.8&15.4&
              13.9&15.1&10.8\\
           $1i_{13/2}$&8.18&8.65&
              7.98&8.84&9.0\\
            $2s_{1/2}$&33.3(23.7)&33.7(24.4)&
              36.2(26.1)&38.0(27.8)&\\
            $2p_{3/2}$&25.8(16.5)&26.1(17.0)&
              27.8(18.1)&28.9(19.1)&\\
            $2p_{1/2}$&25.6(16.3)&25.9(16.8)&
              27.4(17.6)&28.3(18.5)&\\
            $2d_{5/2}$&18.1(8.87)&18.1(8.98)&
              19.4(9.92)&19.5(10.0)&(9.7)\\
            $2d_{3/2}$&17.7(8.50&17.7(8.57)&
              18.5(9.06)&18.6(9.06)&(8.4)\\
            $2f_{7/2}$&10.2&9.80&
              11.0&10.3&9.7\\
            $2f_{5/2}$&9.69&9.25&
              9.85&8.99&7.9\\
            $3s_{1/2}$&16.3(6.75)&16.1(6.63)&
              17.6(7.81)&17.4(7.49)&(8.0)\\
            $3p_{3/2}$&7.88&7.31&
              8.83&7.77&8.3\\
            $3p_{1/2}$&7.69&7.11&
              8.38&7.27&7.4
      \end{tabular}
      \caption{Energy spectra for the $^{208}$Pb nucleus. Energies are in MeV.
      Values between parenthesis are for protons, the others are for neutrons.}
      \label{t5}
\end{table}

\begin{figure}
\epsffile{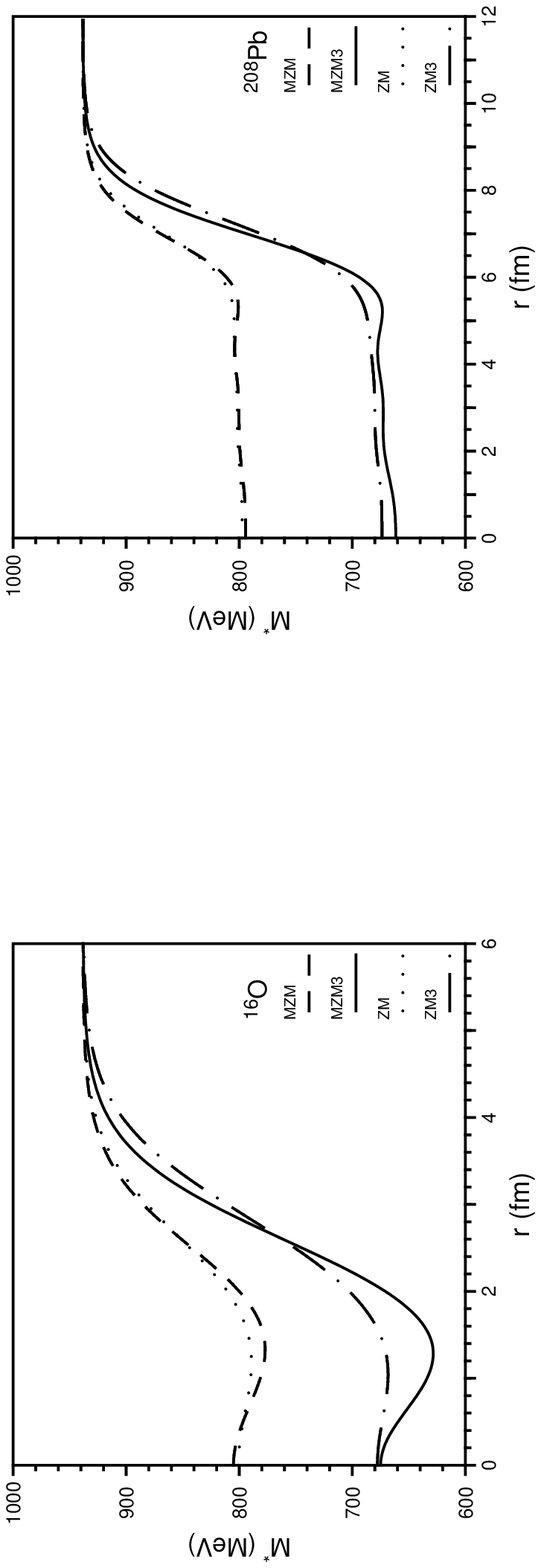}
\caption{The local effective masses $M^\ast$ for $^{16}$O and $^{208}$Pb nuclei.
Besides the curves for the models MZM-MZM3 presented in this work, the
usual ZM-ZM3 ones are shown.}
\label{f3}
\end{figure}

\end{document}